\documentclass[twocolumn,showpacs,reprint,aps,preprintnumbers,superscriptaddress,prc]{revtex4-1}
\usepackage{graphicx}% Include figure files
\usepackage{dcolumn}% Align table columns on decimal point
\usepackage{bm}% bold math
\usepackage{amsmath,amssymb,amsmath}
\usepackage{color}

\begin{document}
\newcommand{\fmi}{\ensuremath{\,\text{fm}^{-1}}}
\newcommand{\mev}{\ensuremath{\,\text{MeV}}}
\newcommand{\kev}{\ensuremath{\,\text{keV}}}
\newcommand{\hw}{\ensuremath{\hbar\omega}}
\preprint{APS/123-QED}

\title{Nuclear structure studies of $^{24}$F \\}

\author{L.~C\'aceres}
\author{A.~Lepailleur}
\affiliation{Grand Acc\'el\'erateur National d'Ions Lourds (GANIL),
CEA/DSM-CNRS/IN2P3, Caen, France}

\author{O.~Sorlin}
\affiliation{Grand Acc\'el\'erateur National d'Ions Lourds (GANIL),
CEA/DSM-CNRS/IN2P3, Caen, France}

\author{M.~Stanoiu}
\affiliation{Grand Acc\'el\'erateur National d'Ions Lourds (GANIL),
CEA/DSM-CNRS/IN2P3, Caen, France}
\affiliation{Institut de Physique Nucl\'eaire, IN2P3-CNRS, F-91406 Orsay
Cedex, France}
\affiliation{Institute of Physics and Nuclear Engineering IFIN-HH, P.O. Box
MG-6, 077125 Bucharest-Magurele, Romania}

\author{D.~Sohler}
\author{Zs.~Dombr\'adi}
\affiliation{Institute for Nuclear Research (MTA Atomki), P.O. Box 51, H-4001 Debrecen, Pf.51, Hungary}

\author{S.~K.~Bogner}
\affiliation{National Superconducting Cyclotron Laboratory and Department of
Physics and Astronomy, Michigan State University, East Lansing, MI 48824, USA}

\author{B.~A.~Brown}
\affiliation{National Superconducting Cyclotron Laboratory and Department of
Physics and Astronomy, Michigan State University, East Lansing, MI 48824, USA}

\author{H.~Hergert}
\affiliation{National Superconducting Cyclotron Laboratory and Department of
Physics and Astronomy, Michigan State University, East Lansing, MI 48824, USA}

\author{J.~D.~Holt}
\affiliation{TRIUMF, 4004 Wesbrook Mall, Vancouver, British Columbia, V6T 2A3 Canada}
\affiliation{Institut f\"ur Kernphysik, Technische Universit\"at
Darmstadt, 64289 Darmstadt, Germany}
\affiliation{ExtreMe Matter Institute EMMI, GSI Helmholtzzentrum f\"ur
Schwerionenforschung GmbH, 64291 Darmstadt, Germany}
\affiliation{National Superconducting Cyclotron Laboratory and Department of
Physics and Astronomy, Michigan State University, East Lansing, MI 48824, USA}

\author{A.~Schwenk}
\affiliation{Institut f\"ur Kernphysik, Technische Universit\"at
Darmstadt, 64289 Darmstadt, Germany}
\affiliation{ExtreMe Matter Institute EMMI, GSI Helmholtzzentrum f\"ur
Schwerionenforschung GmbH, 64291 Darmstadt, Germany}

\author{F.~Azaiez}
\affiliation{Institut de Physique Nucl\'eaire, IN2P3-CNRS, F-91406 Orsay
Cedex, France}

\author{B.~Bastin}
\affiliation{Grand Acc\'el\'erateur National d'Ions Lourds (GANIL),
CEA/DSM-CNRS/IN2P3, Caen, France}

\author{C.~Borcea}
\author{R.~Borcea}
\affiliation{Institute of Physics and Nuclear Engineering IFIN-HH, P.O. Box
MG-6, 077125 Bucharest-Magurele, Romania}

\author{C.~Bourgeois}
\affiliation{Institut de Physique Nucl\'eaire, IN2P3-CNRS, F-91406 Orsay
Cedex, France}

\author{Z.~Elekes}
\affiliation{National Superconducting Cyclotron Laboratory and Department of
Physics and Astronomy, Michigan State University, East Lansing, MI 48824, USA}

\author{Zs. F\"{u}l\"{o}p}
\affiliation{Institute for Nuclear Research (MTA Atomki), P.O. Box 51, H-4001 Debrecen, Pf.51, Hungary}

\author{S.~Gr\'evy}
\affiliation{Grand Acc\'el\'erateur National d'Ions Lourds (GANIL),
CEA/DSM-CNRS/IN2P3, Caen, France}
\affiliation{Universit\'e Bordeaux 1, CNRS/IN2P3, Centre d'\'Etudes
Nucl\'eaires de Bordeaux Gradignan, UMR 5797, Chemin du Solarium, BP.~120, 33175
Gradignan, France}

\author{L.~Gaudefroy}
\affiliation{CEA, DAM, DIF, F-91297 Arpajon, France}

\author{G.~F.~Grinyer}
\affiliation{Grand Acc\'el\'erateur National d'Ions Lourds (GANIL),
CEA/DSM-CNRS/IN2P3, Caen, France}

\author{D.~Guillemaud-Mueller}
\author{F.~Ibrahim}
\affiliation{Institut de Physique Nucl\'eaire, IN2P3-CNRS, F-91406 Orsay
Cedex, France}

\author{A.~Kerek}
\affiliation{Royal Institute of Technology, Stockholm, Sweden}

\author{A.~Krasznahorkay}
\affiliation{Institute for Nuclear Research (MTA Atomki), P.O. Box 51, H-4001 Debrecen, Pf.51, Hungary}

\author{M.~Lewitowicz}
\affiliation{Grand Acc\'el\'erateur National d'Ions Lourds (GANIL),
CEA/DSM-CNRS/IN2P3, Caen, France}

\author{S.~M.~Lukyanov}
\affiliation{FLNR, JINR, RU-141980 Dubna, Moscow region, Russia}

\author{J.~Mr\'azek}
\affiliation{Nuclear Physics Institute, AS CR, CZ-25068 Rex, Czech
Republic}

\author{F.~Negoita}
\affiliation{Institute of Physics and Nuclear Engineering IFIN-HH, P.O. Box
MG-6, 077125 Bucharest-Magurele, Romania}

\author{F.~de~Oliveira}
\affiliation{Grand Acc\'el\'erateur National d'Ions Lourds (GANIL),
CEA/DSM-CNRS/IN2P3, Caen, France}

\author{Yu.-E.~Penionzhkevich}
\affiliation{FLNR, JINR, RU-141980 Dubna, Moscow region, Russia}

\author{Zs.~Podoly\'ak}
\affiliation{University of Surrey, GU2 7XH Guildford, United Kingdom}

\author{M.~G.~Porquet}
\affiliation{CSNSM, CNRS/IN2P3 and Universit\'e Paris-Sud, B\^at 104-108,
F-91405 Orsay, France}

\author{F.~Rotaru}
\affiliation{Institute of Physics and Nuclear Engineering IFIN-HH, P.O. Box
MG-6, 077125 Bucharest-Magurele, Romania}

\author{P.~Roussel-Chomaz}
\author{M.~G.~Saint-Laurent}
\author{H.~Savajols}
\affiliation{Grand Acc\'el\'erateur National d'Ions Lourds (GANIL),
CEA/DSM-CNRS/IN2P3, Caen, France}

\author{G.~Sletten}
\affiliation{Niels Bohr Institute, University of Copenhagen, Denmark}

\author{J.~C.~Thomas}
\affiliation{Grand Acc\'el\'erateur National d'Ions Lourds (GANIL),
CEA/DSM-CNRS/IN2P3, Caen, France}

\author{J.~Tim\`{a}r}
\affiliation{Institute for Nuclear Research (MTA Atomki), P.O. Box 51, H-4001 Debrecen, Pf.51, Hungary}

\author{C.~Timis}
\affiliation{Institute of Physics and Nuclear Engineering IFIN-HH, P.O. Box
MG-6, 077125 Bucharest-Magurele, Romania}

\author{Zs.~Vajta}
\affiliation{Institute for Nuclear Research (MTA Atomki), P.O. Box 51, H-4001 Debrecen, Pf.51, Hungary}

\date{\today}

\begin{abstract}

The structure of the $^{24}$F nucleus has been studied at GANIL using the $\beta$ decay of 
$^{24}$O and the in-beam $\gamma$-ray spectroscopy from the fragmentation of projectile nuclei. 
Combining these complementary experimental techniques, the level scheme of $^{24}$F has been
constructed up to $3.6\mev$ by means of particle-$\gamma$ and particle-$\gamma\gamma$ 
coincidence relations. Experimental results are compared to shell-model calculations 
using the standard USDA and USDB interactions as well as ab-initio valence-space Hamiltonians 
calculated from the in-medium similarity renormalization group based on chiral two- and 
three-nucleon forces.  Both methods reproduce the measured level spacings well, and this close 
agreement allows unidentified spins and parities to be consistently assigned.

\end{abstract}


\maketitle

\section{Introduction}

Nuclear forces play a decisive role in our understanding of the structure of atomic nuclei, driving the 
creation and evolution of shell gaps, the onset of deformation, development of halo structures, 
and determining the limits of particle stability. The nuclear shell model provides a framework to determine 
the properties of nuclei from a set of single-particle energies (SPEs) and two-body matrix elements 
(TBMEs) defined in a given valence space outside some assumed inert core. When based only on
two-nucleon (NN) forces within the valence space, the SPEs and TBMEs need suitable 
renormalization to experimental data to provide a precise description of nuclear structure. These 
``effective'' SPEs and TBMEs implicitly capture the effects of many-body processes, such as core 
polarization, as well as neglected three-nucleon (3N) forces \cite{Caurier,MBPT,Otsu10,Holt12} and 
coupling to the particle continuum.

As the standard shell-model approach typically uses TBMEs that are independent of mass number $A$
or simply scaled, it may 
become insufficient near the limits of stability, where the last nucleons are only loosely bound and
have radial wavefunctions which extend to larger radii and couple to unbound states \cite{Doba07}. 
Reductions of the neutron-neutron TBME by 25\% were required to model the structure of the 
neutron-rich C isotopes, which are a factor of two less bound than the O isotopes 
\cite{Stan08,Camp06}.  More recently, the study of the weakly bound $^{26}$F nucleus, which can be viewed 
as an $^{24}$O core plus a deeply bound $d_{5/2}$ proton and an unbound $d_{3/2}$ neutron, has 
shown that a reduction of the $\pi d_{5/2} - \nu d_{3/2}$ TBME by about 20\% better reproduced the 
energies of its $J=1^+$, $2^+$, $4^+$ states~\cite{Alex}. In addition, recent theoretical calculations 
for states close to the neutron-separation threshold show that an increased role of coupling to the 
particle continuum may in part account for the modification of shell structure of dripline nuclei
\cite{Hamamoto2012,Hagen2012b}. 

The fluorine isotopic chain is well suited to study the evolution of valence-space interactions with increased 
valence neutron-to-proton asymmetry towards the dripline. Moreover, many F isotopes are located 
near the doubly magic $^{16}$O, $^{22}$O and $^{24}$O systems. Considering these O isotopes as 
almost inert cores, fluorine wavefunctions should then be weakly mixed, making possible the search 
for subtle effects related to their weak binding and proximity to the continuum. Located between 
$^{22}$O and $^{24}$O, $^{24}$F is an excellent candidate for such a study, because its spectroscopy is 
expected to be relatively simple. As in $^{26}$F, states located near the neutron separation 
energy, $S_n=3.840(10)\mev$, could be influenced by effects arising from asymmetric 
proton-to-neutron binding.

In a simple shell-model picture, the lowest-lying states in $^{24}$F with spin-parity $2^+$ and $3^+$ can 
be considered pure $\pi d_{5/2} \otimes \nu s_{1/2}$ configurations on top of an $^{22}$O core, while $0^+$ and $1^+$ states are expected from $\pi s_{1/2} \otimes \nu s_{1/2}$ configurations. Close
to the dripline, the excitation of one neutron to the $d_{3/2}$ orbits gives rise to the I$^\pi=1^+-4^+$ 
multiplet due to the $\pi d_{5/2} \otimes \nu d_{3/2}$ coupling observed at low energy in $^{26}$F
\cite{Alex}. Configurations originating from neutron core excitations are also found
below the neutron-separation threshold.

While previous $\beta$-decay studies \cite{mueller,reed} agreed on the $^{24}$O half-life (about 
65~ms), deduced delayed-neutron emission probabilities differed significantly: P$_n=58(12)\%$ in
Ref.~\cite{mueller} and P$_n=18(6)\%$ in Ref.~\cite{reed}. From the latter work, it was expected that 
82(6)\% of the $\beta$ strength would decay to bound states in $^{24}$F~\cite{reed}. Because only 60\% of 
the $\beta$-decay strength was observed, it was proposed that this missing strength may feed 
higher-lying 1$^+$ excited state(s), likely of $\pi d_{5/2}\otimes\nu d_{3/2}$ origin, that could not 
be observed experimentally due to the lack of statistics. Three $\gamma$ transitions associated with 
the decay of $^{24}$O were observed~~\cite{reed}, but the statistics were not sufficient to unambiguously 
establish a $^{24}$F level scheme. No other spectroscopic information was known on $^{24}$F before the 
present study. 

To search for members of the $d_{5/2} - d_{3/2}$ multiplet, the spectroscopy of $^{24}$F 
has been studied using two complementary experimental methods in this work. First, 1$^+$ states were accessed 
from the $\beta$ decay of $^{24}$O, which has an $I^\pi=0^+$ ground state. In a separate experiment, higher-spin states were 
produced in the fragmentation of projectile nuclei leading to $^{24}$F. The results were then compared to 
shell-model calculations based on the benchmark USDA and USDB empirical Hamiltonians 
\cite{USDA} as well as ab initio valence-space Hamiltonians derived from NN+3N forces 
\cite{Bogn14}.  The excellent agreement between these two calculations results in a robust 
description of the newly measured states.

\section{$\beta$-decay of $^{24}$O }
\subsection{Experimental set-up}

\begin{figure}
\includegraphics[width=0.45\textwidth]{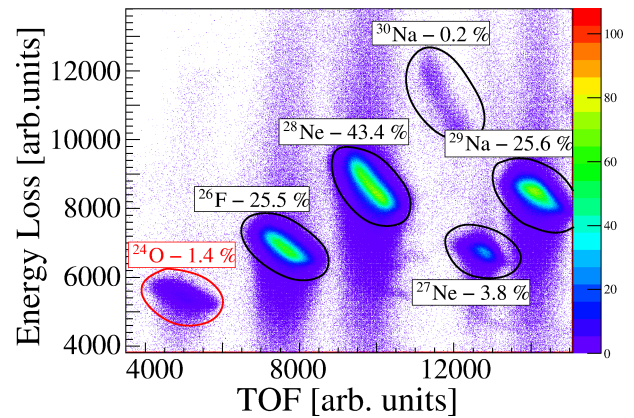}
\caption{\label{fig:ID} (Color online) Energy loss versus time-of-flight
identification matrix. The relative yields of the implanted ions are shown by
different shades.}
\end{figure}

The $^{24}$O nucleus was produced via fragmentation of a $^{36}$S$^{16+}$
primary beam delivered by the GANIL facility in a 237~mg/cm$^2$ Be target placed
at the entrance of the LISE spectrometer~\cite{Anne}. The energy and the average
intensity of the primary beam were 77.6~MeV/u and 2~e$\mu$A, respectively. The
projectile-like fragments were separated by the LISE achromatic spectrometer. A
$^9$Be wedge-shaped degrader of 1066~mg/cm$^2$ was placed at the dispersive
focal plane of LISE to improve the ion selection. As shown in Fig.~\ref{fig:ID} the
selected nuclei were identified at the end of the spectrometer by means of their
energy loss ($\Delta$E) in two silicon detectors of 500~$\mu$m thickness and their
time-of-flight (TOF) referenced to the cyclotron radio-frequency. An Al foil of
adjustable inclination was placed after the two Si detectors to allow the
implantation depth of the  $^{24}$O ions into a 1~mm
double-sided-silicon-strip detector (DSSSD) of 5$\times$5~cm$^2$ with
16$\times$16 strips. The pixels in the DSSSD were used to establish spatial
and time correlations between the $\beta$-particles and the $^{24}$O parent. Energy threshold of the individual strips were set to
$\sim$80~keV. A 5~mm-thick Si detector was placed after the DSSSD to control the
implantation depth of $^{24}$O. 

Four segmented Ge Clover detectors of the EXOGAM array~\cite{EXO} were placed
around the DSSSD detector to provide $\beta-\gamma$ coincidences. A
$\gamma$-ray efficiency $\varepsilon_{\gamma}$ of 6.5\% at 1 MeV was extracted from the
$\beta$-decay of $^{28}$Ne which was transmitted in the same set-up and for
which the intensities of the $\gamma$ transitions were known ~\cite{28Ne}. For
each implanted nucleus, a $\beta$ efficiency ($\varepsilon_{\beta}$) of 63(3)\% was
extracted from the intensity ratio between an identified $\gamma$-line gated or ungated on the $\beta$-correlation condition. The relative $\gamma$-ray intensities
were obtained from the $\varepsilon_{\beta}$ and $\varepsilon_{\gamma}$ values. 
\subsection{Results}

\begin{figure}
\includegraphics[width=0.5\textwidth]{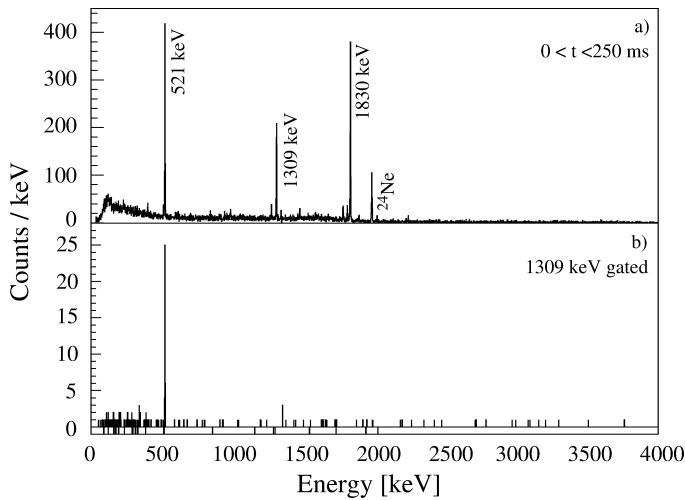}
\caption{\label{fig:beta} a) $\beta$-gated $\gamma$-ray spectrum
following the implantation of $^{24}$O  nuclei obtained in the 0--250~ms range.
b) $\gamma\gamma$-coincidence spectrum gated by the 1309~keV $\gamma$-ray. }
\end{figure}

The $\beta$-gated $\gamma$-ray spectrum following the implantation of a
precursor $^{24}$O nucleus between 0--250~ms is shown in Fig.~\ref{fig:beta} a).
This time condition favors the observation of $\gamma$ rays associated to the
decay of $^{24}$F while it suppresses all transitions belonging to
contaminant nuclei or daughter activity. The previously reported
$\gamma$-lines~\cite{reed} at 521~keV, 1309~keV and 1830~keV are clearly
visible. All other observed transitions are not attributed to the decay of $^{24}$O. 

A total of $\sim$10$^5$ $^{24}$O nuclei were implanted in the DSSSD detector, which is a factor of 10 larger than
in Ref.~\cite{reed}. The improved statistics obtained in the present work permitted a study of $\gamma\gamma$
coincidences (Fig.~\ref{fig:beta} b), from which it was deduced that the 1$^+$
level at 1830~keV is fed with $I_\beta$=57(4)~\% and decays by a cascade of
521~keV and 1309~keV
$\gamma$-rays. The ordering of these two transitions could not be determined unambiguously
from the $\beta$-decay study as they have the same relative intensity
(Table~\ref{tab:intenB}). The in-beam experiment presented in
Section III~B shows that the 1309~keV $\gamma$-ray feeds a level at 521~keV. The
1830~keV level decays by a competing branch directly to the ground state, as
well. The $\gamma$-ray energies, relative intensities and branching ratios of
the observed transitions are listed in Table~\ref{tab:intenB}.

\begin{figure}
\includegraphics[width=0.45\textwidth]{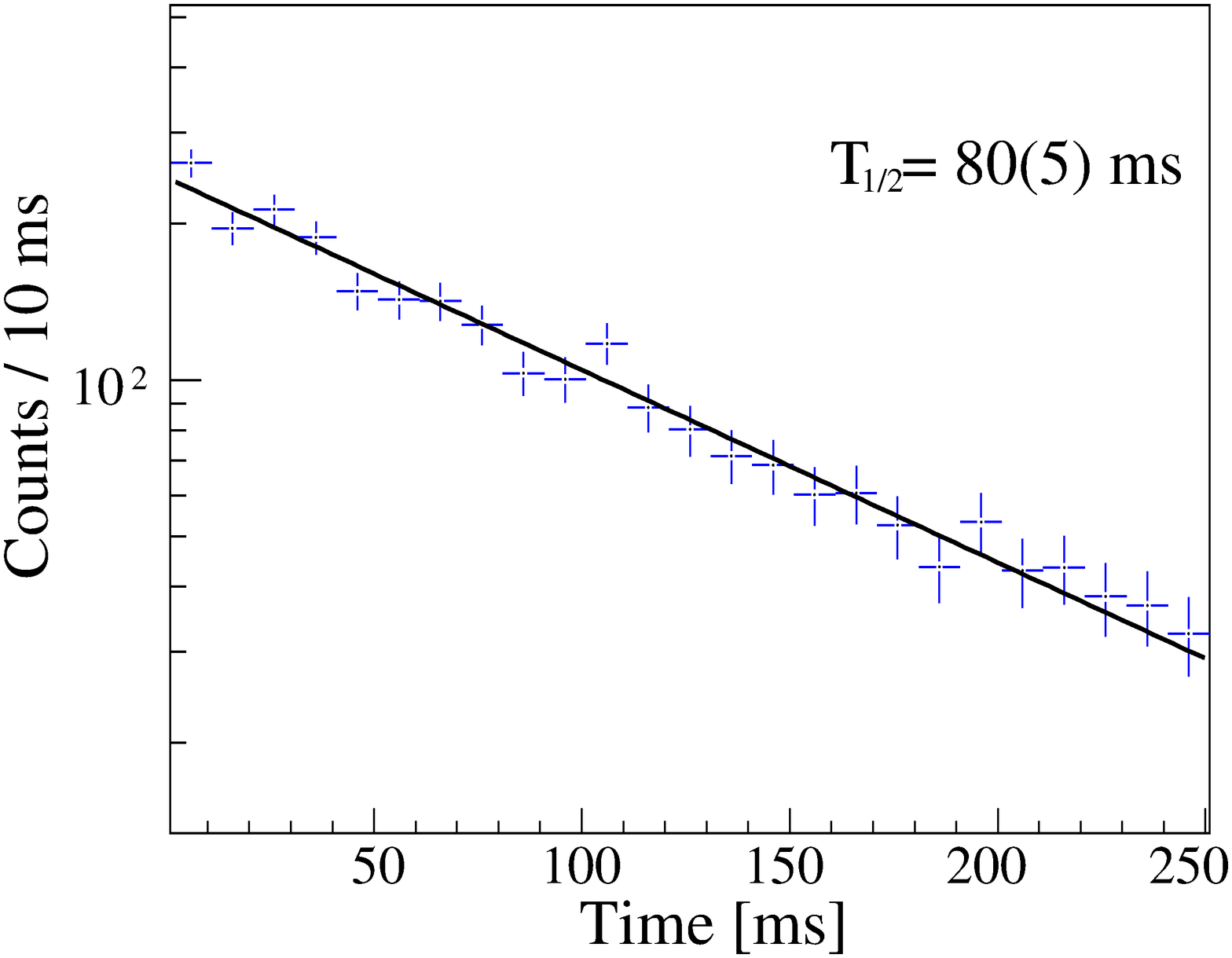}
\caption{\label{fig:time} (Color online) Background-subtracted summed
time-distribution gated on the 521~keV, 1309~keV and 1830~keV $\gamma$-rays 
following the implantation of $^{24}$O nuclei. The fit with a single exponential
decay-curve yields T$_{1/2}$=80(5)~ms.}
\end{figure}

The background subtracted summed time-distribution of the 521~keV, 1309~keV and
1830~keV $\gamma$-rays with respect to the $^{24}$O implantation is shown in
Fig.~\ref{fig:time}. A fit with a single exponential decay-curve yields a
half-life of T$_{1/2}$=80(5)~ms, in agreement with the 61$^{+32}_{-19}$~ms value of Ref.~\cite{mueller}
within the statistical uncertainties but longer than the value of 65(5)~ms
reported in Ref.~\cite{reed}. Using the $\beta$-decay Q-value and the absolute
feeding intensity of the 1830~keV state a log$ft$ value of 4.25(6) has been
obtained, which is consistent with an allowed Gamow-Teller transition. 

The observational limit for the population of other 1$^+$ states has been
measured to be 1.0(4)\%. The $\beta$-delayed neutron emission probability has
been extracted to be $P_n$=43(4)\% from the yields of the 1830~keV and 1309~keV
transitions normalized to the total number of $^{24}$O detected. This value is
in agreement with P$_n$~=~58(12)\% previously reported in Ref.~\cite{mueller}
but larger than the value of 18(6)\% measured in Ref.~\cite{reed}. 

In a separate setting of the LISE spectrometer, the $\beta$-decay of $^{24}$F
was studied as well. From this data set the direct feeding to the  2$^+$ state
at 1981~keV and the 4$^+$ state at 3963~keV in $^{24}$Ne were observed~\cite{alex2}. 
It establishes that the spin and parity of the ground state of $^{24}$F is 3$^+$.
It follows that the 521~keV state is located in between the 3$^+$ ground state
and the 1$^+$ excited state at 1830~keV. As the 521~keV state is not populated
directly in the $^{24}$O $\beta$-decay its spin and parity must therefore be 2$^+$. 

\begin{table}
\caption{\label{tab:intenB} Experimental energies, spin and parity assignments,
transition energies, relative intensities per 100 decays. $\gamma$ branching
ratio (BR) for the excited states of $^{24}$F observed in the $\beta$-decay of
$^{24}$O experiment.}
%\begin{minipage}{10.3cm}
 \begin{tabular}{cccccccc}
\hline
\hline
E$_i$ [keV] &I${_i^\pi} \rightarrow$ I${_f^\pi}$&E$_\gamma$ [keV]&
I$_\gamma$[$\backslash$ 100 decays]&BR[$\%$]\\
\hline
521(1)&2$^+_1\rightarrow$3$^+_1$&521(1)&21(2)&\\
1830(1)&1$^+_1\rightarrow$3$^+_1$&1830(1)&39(3)&68(5)\\
&1$^+_1\rightarrow$2$^+_1$&1309(1)&18(2)&32(3)\\
\hline
\hline
\end{tabular}
 %\end{minipage}
\end{table}

\section{In-beam spectroscopy of the $^{24}$F nucleus}
\subsection{Experimental set-up}
The $^{24}$F nucleus was produced in a two-step reaction. A $^{36}$S primary
beam of 77.5~A~MeV with an average intensity of 6.5~e$\mu$A underwent fragmentation in a
398~mg/cm$^2$ C target placed between the two superconducting solenoids of the
SISSI~\cite{sissi} device. The reaction products were separated and selected
trough the ALPHA spectrometer by means of the B$\rho$-$\Delta$E-B$\rho$
method~\cite{dufour,Anne}. The transmitted cocktail beam was composed of
$^{25,26}$Ne, $^{27,28}$Na and $^{29,30}$Mg nuclei with energies ranging between 54
and 65~MeV/u. The identification of the beam ions was performed by the combined
measurement of their time-of-flight (TOF) over a flight path of 80~m using two
micro-channel plates and their energy loss in a plastic scintillator of
103.5~mg/cm$^2$ thickness located at the entrance of the SPEG spectrometer~\cite{SPEG}. Two
C foils of 51~mg/cm$^2$ thickness were placed before and after the plastic
scintillator constituting a secondary ``active'' target. The $^{24}$F nuclei
were produced in this secondary target through the fragmentation of $^{27}$Na.
Once produced, the $^{24}$F nuclei were separated from other reaction residues
in the SPEG spectrometer. The identification of the ions was performed on an
event-by-event basis at the final focal plane of the SPEG by measuring their
time-of-flight (TOF) with a plastic scintillator, their energy loss ($\Delta$E)
and position in an ionization and two drift chambers, respectively. In addition,
74~BaF$_2$ detectors of the Ch\^{a}teau de Cristal array surrounded the
secondary target at an average distance of 30~cm. Prompt $\gamma$-ray emission
was measured in coincidence with the nuclei identified at the final focal plane
of SPEG. The photopeak efficiency of the array was 24~\%, 42~\% and 29~\% for
$\gamma$-ray energies of 100~keV, 600~keV and 1300~keV, respectively. 

\subsection{Results}

\begin{figure}
\includegraphics[width=0.45\textwidth]{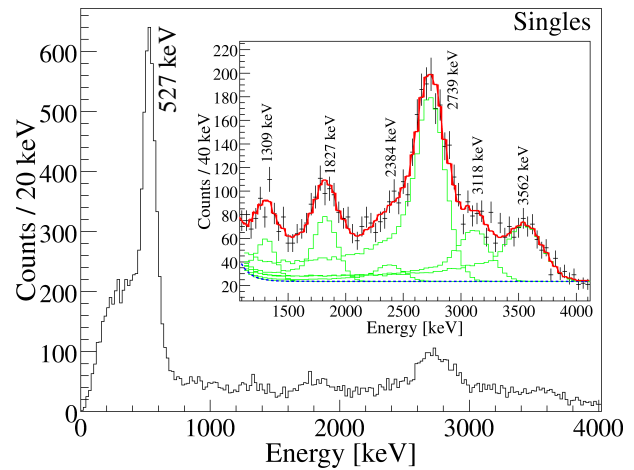}
\caption{\label{fig:HE} (Color online) Singles $\gamma$-ray spectrum obtained in
coincidence with the $^{24}$F nuclei produced in the in-beam $\gamma$-ray
spectroscopy experiment. The inset presents a zoom on the
high-energy part of the spectrum. The lines show the result of the fit with line
shapes obtained from {\sc Geant} simulation.}
\end{figure}

\begin{figure}
\includegraphics[width=0.45\textwidth]{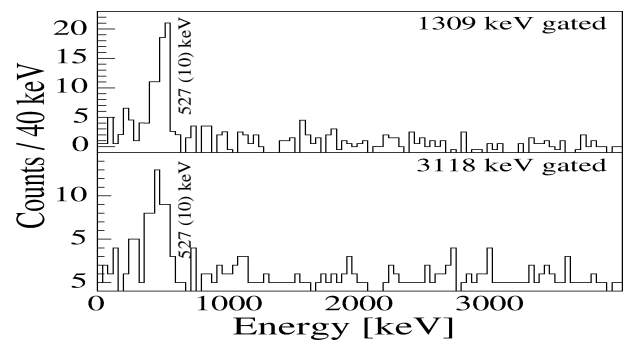}
\caption{\label{fig:coin} $\gamma\gamma$ coincidence-spectra gated by the
1309~keV (a) and 3118~keV (b) transitions.}
\end{figure}

Prompt $\gamma$ rays observed in coincidence with the $^{24}$F nuclei identified in SPEG are listed in Table~\ref{tab:inten}. The singles
$\gamma$-ray spectrum of $^{24}$F is shown in Fig.~\ref{fig:HE}. The three
$\gamma$ rays at 527(10)~keV, 1309(22)~keV and 1827(11)~keV correspond, within
the experimental uncertainties, to those observed in the $\beta$-decay of
$^{24}$O at 521(1)~keV, 1309(1)~keV and 1830(1)~keV, respectively. 
The excitation energy of the 1829(26)~keV state has been extracted by the
weighted mean of the energies of the two decay branches. It is in agreement with
the value obtained in the $\beta$-decay data set. 
The $\gamma\gamma$-coincidence between the 527~keV and 1309~keV transitions
(Fig.~\ref{fig:coin} a) is confirmed. The ordering of the two transitions is
obtained from their relative intensity since the 527~keV $\gamma$-ray intensity is much
larger than that of the 1309~keV transition (Tab.~\ref{tab:inten}) the
527~keV state is placed below. It can be fed directly in the reaction and/or by
other higher-lying excited states. In addition, four new transitions have been
observed between 2200~keV and 4000~keV. To get a reasonable line shape, the
response function of the BaF$_{2}$ array was simulated using the
\textsc{Geant4} package. In the simulation the energy dependence of the peak
width, the cut-off energy, the Doppler shift and the Doppler broadening were
taken into account.  The energy-dependent width of the $\gamma$ ray peak has been
extracted from the spectroscopy of other nuclei produced in similar experimental
conditions~\cite{thesis}. For $\gamma$-rays with energies greater than
$\sim$1.5~MeV, in addition to the photo-absorption and Compton effects pair
creation also starts to play a significant role. Since an add-back procedure was
used in the analysis of the $\gamma$-ray spectra, the escape of the annihilation and
Compton scattered $\gamma$ rays are suppressed. As a consequence, the line shape
is well described using a Gaussian plus a long, low-energy tail. Using the simulated line shapes
the $\gamma$-ray spectrum between 2.2 and 4.0~MeV could be described by four
$\gamma$-rays at 2384(64), 2739(14), 3118(33) and 3562(22)~keV (inset of
Fig.~\ref{fig:HE}). 

A clear coincidence between the 527~keV  and 3118~keV $\gamma$-rays is observed
(Fig.~\ref{fig:coin} b)), establishing a state at 3639~keV excitation energy, considering that the 527~keV transition corresponds to the 521~keV $\gamma$-line observed in the $\beta$-decay experiment. No
other $\gamma\gamma$-coincidences were observed, indicating that the
2384~keV, 2739~keV and 3562~keV transitions decay directly to the ground
state. Therefore, three new excited states are proposed at 2384(64)~keV,
2739(14)~keV and 3562(22)~keV.  

\section{Discussion}

\begin{figure*}
\includegraphics[width=0.95\textwidth]{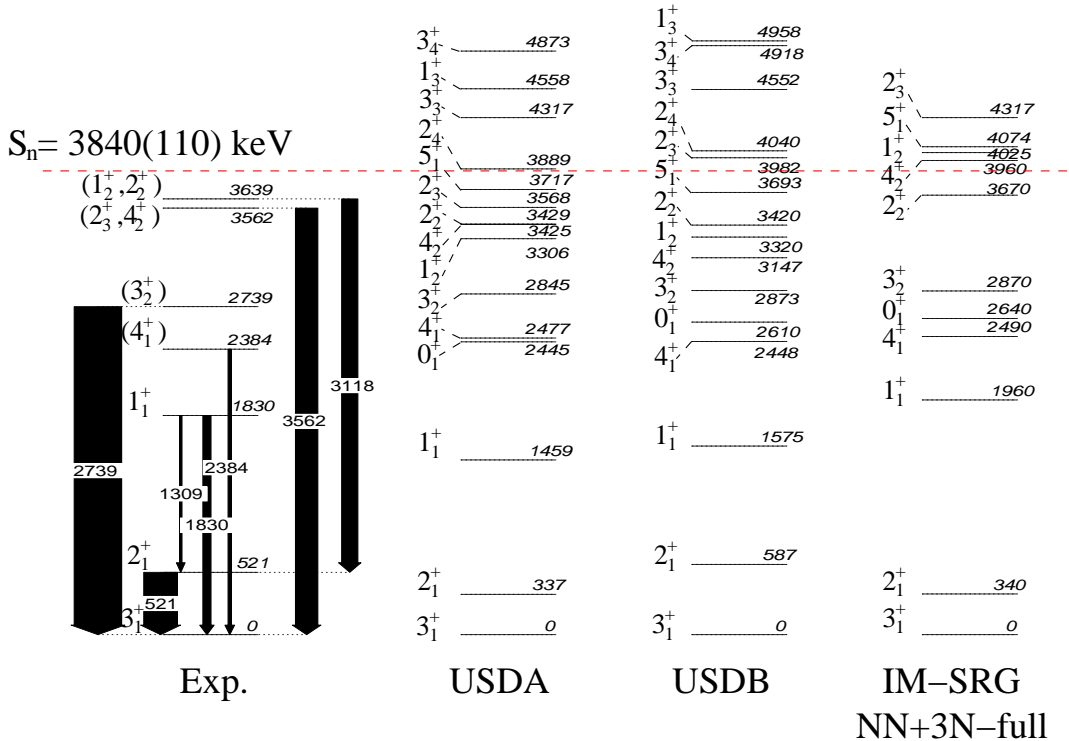}
\caption{\label{fig:scheme} (Color on line) Experimental level scheme of
$^{24}$F obtained in the in-beam and $\beta$-decay data set compared to shell-model calculations
performed with the USDA and USDB interactions as well as ab-initio valence-space Hamiltonians calculated from the in-medium similarity renormalization group.
}
\end{figure*}

The experimental level scheme deduced in the discussed experiments up to the neutron separation energy of 3.84~MeV is shown in
Fig.~\ref{fig:scheme}. The spin and parity of the previously reported 1830~keV
state~\cite{reed} can be firmly established to be 1$^+$ from the $\beta$-decay
of $^{24}$O. The level scheme has been completed by four new states at higher
excitation energy. In order to clarify their spin and parity assignment, the
experimental excitation energies and branching ratios were compared to
predictions of different theoretical calculations (Fig.~\ref{fig:scheme}). 

The shell-model calculations were performed using the standard USDA and USDB
interactions~\cite{USDA, USDAm} in  the full $sd$ valence space. The data used
in the USDA/USDB fit comprises ground states and low-lying excited states of the
$sd$-shell nuclei from A~=~16 to A~=~40, therefore the experimental binding energy of
the ground state of $^{24}$F is in agreement with the theoretical predictions.

In order to assign the spins of the observed states, the experimental branching
ratios were compared to the theoretical ones extracted from the E2 to M1 decay
branch of each state, re-scaled in energy to match the experimental value.
Starting from spin assignments of the low-energy levels, the higher-energy ones
can be deduced. In the calculation of the M1 reduced transition probabilities, the
effective g-factors of the proton (neutron) g$_{sp}$~=~5.0 (g$_{sn}$~=~-3.44),
g$_{lp}$~=~1.174 (g$_{sn}$~=~-0.110) and g$_{tp}$~=~0.24 (g$_{tn}$~=~-0.16) for
the spin, orbital and tensor components of the M1 operator are taken from
Ref.~\cite{USDAm}. The E2 reduced transition probabilities are calculated using effective charges
of 1.36e and 0.45e for protons and neutrons, respectively.

\begin{table}
\caption{\label{tab:inten} Experimental energies, tentative spin and parity
assignments, transition energies, relative feeding intensities and $\gamma$
branching-ratios (BR) for the excited states in $^{24}$F obtained in the in-beam
spectroscopy experiment.}
%\begin{minipage}{10.3cm}
 \begin{tabular}{cccccccc}
\hline
\hline
E$_i$ [keV] &I${_i^\pi}$ $\rightarrow$ I${_f^\pi}$&E$_\gamma$ [keV]&
I$_\gamma[\%]$&BR[\%]&\\
\hline
527(10)&2$^+_1\rightarrow$3$^+_1$&527(10)&71(3)&\\
1829(26)&1$^+_1\rightarrow$3$^+_1$&1827(11)&17(2)&77(10)\\
&1$^+_1\rightarrow$2$^+_1$&1309(22)&5(1)&23(5)\\
2384(64)&(4$^+_1$)$\rightarrow$3$^+_1$&2384(64)&7(3)&\\
2739(14)&(3$^+_2$)$\rightarrow$3$^+_1$&2739(14)&100(5)&\\
3562(22)&(2$^+_3$, 4$^+_2$)$\rightarrow$3$^+_1$&3562(22)&47(5)&\\
3639(42)&(1$^+_2$, 2$^+_2$)$\rightarrow$2$^+_1$&3118(33)&34(3)&\\
\hline
\hline
\end{tabular}
 %\end{minipage}
\end{table}

The spin and parity assignment of the experimental state at 527~keV is 2$^+_1$.
The USDA interaction underestimates the excitation energy by
$\sim$200~keV while better agreement is found in the USDB calculations. This 2$^+_1$ level decays 100\% to the 3$^+$ ground state of $^{24}$F. Both levels belong to the $\pi$d$_{5/2}\otimes\nu$s$_{1/2}$ multiplet
with an almost pure wave function ($\sim$70\%). 

The excitation energy of the
1$^+_1$ state at 1830~keV, which originates mainly ($\sim$50\%) from a  mixed $\pi$s$_{1/2}\otimes\nu$s$_{1/2}$ configuration, is underestimated in the USDA and USDB calculations. A somehow similar shift in energy between experiment and theory is found for the 1/2$^+$ state in $^{25}$F \cite{Vajta}, that is due to an $s_{1/2}$ proton excitation. The 1$^+_1$ level decays by a 1830~keV transition
to the ground state (77(10)\%) and by a parallel branch to the 521~keV level
(23(5)\%), which agrees with the USDB calculations of 72\% and 28\%,
respectively. The USDA calculations result in 95\%
decay to the ground state and 5\% to the 2$^+_1$ level. 
The reduced M1 matrix element is 0.11~$\mu_N$ with USDA and
0.17~$\mu_N$ with USDB. This difference is consistent with
the results shown in Figure 2 of Ref.~\cite{USDAm} where 
M1 matrix elements for USDA and USDB differ in 
a random way with a root-mean square (r.m.s.) difference of about 0.10~$\mu_N$
being about the same for large and small values.
As suggested in~\cite{USDAm}, the r.m.s. difference
between experimental and theoretical M1 matrix
elements might be reduced if some of the M1 data
could be used to constrain the Hamiltonian.

The level at 2384~keV decays exclusively to the ground state. The only theoretical counterpart having
the same decay pattern is the 4$^+_1$ state, which has a 73\% pure 
$\pi$d$_{5/2}\otimes\nu$(d$_{5/2})^{-1}$(s$_{1/2})^2$ configuration.

The energy spacing of the experimental levels between 2400~keV and 3700~keV is
better reproduced in the calculations performed with the USDA interaction. The
results obtained with the USDB interaction show a spectrum with states having a
regular spacing. Both calculations predict a high density of
levels at high excitation energy close to the neutron separation energy, among
them the 1$^+_2$ and 4$^+_2$ states belonging to the
$\pi$d$_{5/2}\otimes\nu$d$_{3/2}$ multiplet.  These levels are of the same
origin as the 1$^+_1$ and 4$^+_1$ states in the low-energy spectrum of the
weakly bound $^{26}$F nucleus. 

%For this nucleus, the authors of Ref.~\cite{Alex} observed a
%compression of the experimental J~=~1$^+-$4$^+$ multiplet and a reduction of the binding
%energy of the J~=~1$^+$,2$^+$,4$^+$ states with respect to the theoretical values
%calculated using the USDA interaction. It was concluded that the residual interaction
%between these high-lying states is weakened by many-body correlations and continuum 
%effects which are not taken into account when USDA/B matrix elements are fit. The
%present findings suggest that the same mechanism is at play in $^{24}F$.

The spin and parity of the 2739~keV state is tentatively assigned to be 3$^+_2$ because any
other possibility would imply a deviation between the experimental and
theoretical excitation energy of more than 400~keV.
Experimentally, this state decays by more than 95\% 
to the 3$^+_1$ ground state in contrast
to the calculations that give 3, 84, 12~\%
to the 3$^+_1$, 2$^+_1$, 4$^+_1$ states for USDA, respectively, and 15, 38, 47~\%
for USDB.
The difference between experiment and
theory might come from the fact that the B(M1) value
to the ground state is very small:
0.0010~$\mu_N^2$ for USDA and 0.0017~$\mu_N^2$ for USDB.
A B(M1) value of 0.010~$\mu_N^2$ would give 90~\% branch to the ground state.
The corresponding M1 matrix elements are
0.083~$\mu_N$ (USDA) and 0.110~$\mu_N$ (USDB) and 0.26~$\mu_N$ for B(M1)=0.010~$\mu_N^2$.
This range of matrix elements is within the overall
best-fit r.m.s. deviation between experimental and theoretical
M1 matrix elements of 0.20~$\mu_N$ (Table I in~\cite{USDAm}).
The 3$^+_2$ level belongs to the same multiplet as the 4$^+_1$ state with a $\sim$53\% pure configuration. 

At higher excitation
energies there is a large density of calculated levels which make the spin
and parity assignments to the experimental 3562~keV and 3639~keV states difficult. The
experimental branching ratios were compared to the calculated decay pattern of the 4$^+_2$,
2$^+_3$, 2$^+_2$ and 1$^+_2$ levels. The I$^\pi$ of the 3562~keV level is
tentatively assigned to be 4$^+_2$ or 2$^+_3$, that of the 3639~keV state 2$^+_2$ or
1$^+_2$. The 3562~keV state is a good candidate for the 4$^+_2$ which belongs to
the $\pi$d$_{5/2}\otimes\nu$d$_{3/2}$ multiplet with $\sim$64\% pure
configuration. It would be favorably populated in the reaction due to being an Yrast state. 
As no evidence of the existence of a $1^+_2$ state was found in the $\beta$-decay 
experiment, the 3639 keV state is likely to be I~=~$2^+_2$. The 2$^+_2$ and 2$^+_3$ levels 
have very mixed and complex wave functions. 

In addition, theoretical predictions from first-principles valence-space 
calculations, where the many-body processes and 3N forces are included are explored.  Effective 
shell-model Hamiltonians based on two-nucleon (NN) and 3N forces were first 
derived for the sd-shell region within the context of many-body perturbation theory 
\cite{Otsu10,Holt13a,Holt13b,Caes13,Gallant}, where excitations outside the valence 
space were calculated to third order. There it was found that both 3N forces as well 
as an extended valence space (i.e., including orbitals beyond the standard sd shell), 
were essential to describe semi-magic isotopic/isotonic chains on top of a $^{16}$O 
core. The need for an extended space in the perturbative calculation of the 
valence-space Hamiltonian suggests that the extended orbitals need to be 
included nonperturbatively.

Therefore, it has been considered in this paper a novel nonperturbative method for constructing 
valence-space Hamiltonians: the in-medium similarity renormalization group (IM-SRG) 
\cite{Tsuk11,Tsuk12,Hergert13,Bogn14}. In addition to the IM-SRG other ab initio 
methods have now successfully treated the oxygen chain and select fluorine and neon 
isotopes with NN+3N forces \cite{CCOx,Herg13,Cipo13,CCGT}. In the 
IM-SRG, a continuous unitary transformation, parameterized by the flow parameter 
$s$, is applied to the initial normal-ordered $A$-body Hamiltonian such that undesirable 
off-diagonal couplings are driven to zero as $s\rightarrow\infty$:
\begin{gather}
H(s)=U^{\dagger}(s)HU(s)=H^{\rm d}(s)+H^{\rm od}(s)\,,\\
H^{\rm od}(s\rightarrow\infty)=0\,.
\end{gather}
Taking the uncorrelated ground state of doubly magic $^{16}$O, and defining $H^{\rm od}$ to be all 
$n$-particle-$n$-hole excitations, $H(s\rightarrow\infty)$ will flow to the fully correlated (i.e., 
exact) ground-state energy as $H^{\rm od}(s)\rightarrow0$. Including excitations that connect
valence-space to non-valence-space particle states in the definition of $H^{\rm od}$, the $sd$ 
valence space will decouple from the core and higher shells as $s\to\infty$. The resulting 
Hamiltonian $H(\infty)$ will then consist of renormalized $sd$-shell SPEs
and TBMEs, to be used as input in a standard shell-model calculation, in addition to the $^{16}$O 
core energy \cite{Tsuk12,Bogn14}.  

The starting point for these calculations are nuclear forces derived from chiral effective field theory 
\cite{EpelRMP,EMPR}. We use the $500\mev$-cutoff N$^3$LO NN potential of Ref.~\cite{EM500} and 
the local N$^2$LO 400~MeV-cutoff 3N interaction of Ref.~\cite{Roth3N400}, evolved with the 
free-space SRG \cite{BognSRG} to a lower momentum scale, $\lambda_{\rm SRG}=1.88\fmi$. 
IM-SRG $sd$-shell Hamiltonians are then calculated following the procedure outlined above, based 
on SRG-evolved NN forces with 3N forces induced by the SRG evolution (NN+3N-induced) as well as 
with initial 3N forces (NN+3N-full). The latter are included through normal ordering with respect to the 
$^{16}$O Hartree-Fock reference state, truncated at the two-body level \cite{CCNO3N,RothNO}.  For 
complete details, see Ref.~\cite{Bogn14}. Finally, the resulting shell-model Hamiltonians are 
diagonalized to obtain the spectrum of $^{24}$F, studying for the first time IM-SRG proton-neutron 
valence interactions. 

Without initial 3N forces, the spectrum (not shown) is much too compressed and the ordering of levels
is incorrect: The first eight excited states lie below $2.0\mev$, in clear contrast to experiment and the 
NN+3N-full results shown in Fig.~\ref{fig:scheme}.  While the ground-state energy of $^{24}$F is 
overbound by 7.7~MeV in the NN+3N-full calculation, the predicted excited-state spectrum is in 
remarkably good agreement with the new experimental measurements.  In particular, all excited 
states below the one-neutron separation threshold are less than 200~keV away from corresponding 
experimental levels. The only exception is a $0^+_1$ state at $2640\kev$, also predicted with USDA,B, 
which is likely not seen experimentally due to the difficulty of the fragmentation method in populating 
low-$J$ states.  Since coupling to the continuum is currently neglected, when included, a modest 
lowering of the $2^+_2$, $4^+_2$, $1^+_1$, and $5^+_1$ states would be expected near threshold.  
Furthermore the wavefunctions and $\gamma$ transitions involving these states are very similar to 
those discussed above for USDA,B, strengthening the proposed identifications made for the $3^+_2$ 
and $4^+_1$ states. Specifically the $3^+_2\rightarrow3^+_1$ branch is 26\%, in moderately better agreement 
with experiment.  The probable
identification of the $3639\kev$ level as the $2^+_2$ state suggests an incorrect $2^+_2 - 4^+_2$ 
ordering is seen in these calculations. This can be understood in terms of neglected continuum 
effects: The $\nu$d$_{3/2}$ component of the 4$^+_2$ state is twice as large as that in 2$^+_2$, hence
one would naively expect that, when added, the continuum would then lower the 4$^+_2$ by a greater amount than the 2$^+_2$, possibly resulting in the correct ordering. This hypothesis needs to be confirmed by an unambiguous location of the $2^+_2$ and $4^+_2$ states in $^{24}$F and more detailed theoretical calculations.
Recent coupled-cluster calculations 
based on optimized N$^2$LO chiral NN and 3N forces have been performed for $^{24}$F 
\cite{CCGT}, which exhibit reasonable agreement with this new experimental picture.

\section{Summary}

Detailed spectroscopy of the $^{24}$F nucleus has been obtained at GANIL using two complementary 
experimental techniques: $\beta$ decay and in-beam $\gamma$-ray spectroscopy from projectile fragmentation. 
Previously reported transitions (521, 1309 and 1830~keV) have been confirmed, and in addition four 
new $\gamma$ rays have been observed for the first time. The $\gamma$-ray ordering was established 
from relative intensity arguments and the large statistics of the present data allowed to perform 
a $\gamma\gamma$-coincidences analysis. Gathering all the available information on $^{24}$F, a level 
scheme has been proposed up to the neutron separation energy. The ground state spin and parity of 
$^{24}$F is unambiguously determined to be 3$^+$. Excitation energies and branching ratios are 
compared to two shell-model calculations (using the standard USDA and USDB interactions) as well
as to ab initio shell-model calculations, using interactions derived from chiral NN+3N forces 
by means of the IM-SRG. 
From this comparison a clear identification of almost all measured states has been obtained. 
It is suggested that the  3$^+_1$ ground state and the 2$^+_1$ levels  belong to the 
$\pi$d$_{5/2}\otimes\nu$s$_{1/2}$ multiplet with more than 70\% pure wavefunctions.  The 
 1$^+_1$ state has a predominant  $\pi$s$_{1/2}\otimes\nu$s$_{1/2}$ configuration, while the 4$^+_1$ and 3$^+_2$ states have predominant $\pi$d$_{5/2}\otimes\nu$(d$_{5/2})^{-1}$(s$_{1/2})^2$ configurations. At higher excitation energy, the large density of observed and calculated states 
makes the identification of the experimental levels more ambiguous. Tentative $2^+_2$ and 4$^+_2$ spin parity values are proposed for the $3639\kev$ and 3562~keV levels, respectively.  Their ordering is only reproduced by the shell model using the USDA 
and USDB interactions, while the calculations performed with the IM-SRG predicts an
inversion of the two states. As being significantly less mixed than the 2$^+_2$ state, the 4$^+_2$ is expected to be more sensitive to continuum effects, and therefore the explicit treatment of such effects should lower its energy by a 
greater amount than the 2$^+_2$. This would possibly result in the correct ordering in the IM-SRG model. Further spectroscopic information of $^{24}$F at high excitation energy and more detailed theoretical calculations will be needed to elucidate 
the role of the continuum in the high-energy part of the level scheme of this nucleus, but the overall agreement between IM-SRG theory and experiment is globally extremely satisfactory.

\begin{acknowledgments}
The authors are thankful to the GANIL and LPC staffs and the EXOGAM collaboration. 
This work has been supported by the European Community contract no. RII3-CT-2004-506065, 
by OTKA K100835 and NN104543, by the NSF PHY-1068217 grants, by a grant of the 
Romanian National Authority for Scientific Research, CNCS - UEFISCDI, project no. 
PN-II-ID-PCE-2011-3-0487. We thank S.~Binder, A.~Calci, 
J.~Langhammer, and R.~Roth for providing us with chiral 3N matrix elements. Computing 
resources for the IM-SRG calculations were provided by the Ohio Supercomputer Center (OSC).
\end{acknowledgments}


\begin{thebibliography}{0}%
\makeatletter
\providecommand \@ifxundefined [1]{%
 \@ifx{#1\undefined}
}%
\providecommand \@ifnum [1]{%
 \ifnum #1\expandafter \@firstoftwo
 \else \expandafter \@secondoftwo
 \fi
}%
\providecommand \@ifx [1]{%
 \ifx #1\expandafter \@firstoftwo
 \else \expandafter \@secondoftwo
 \fi
}%
\providecommand \natexlab [1]{#1}%
\providecommand \enquote  [1]{``#1''}%
\providecommand \bibnamefont  [1]{#1}%
\providecommand \bibfnamefont [1]{#1}%
\providecommand \citenamefont [1]{#1}%
\providecommand \href@noop [0]{\@secondoftwo}%
\providecommand \href [0]{\begingroup \@sanitize@url \@href}%
\providecommand \@href[1]{\@@startlink{#1}\@@href}%
\providecommand \@@href[1]{\endgroup#1\@@endlink}%
\providecommand \@sanitize@url [0]{\catcode `\\12\catcode `\$12\catcode
  `\&12\catcode `\#12\catcode `\^12\catcode `\_12\catcode `\%12\relax}%
\providecommand \@@startlink[1]{}%
\providecommand \@@endlink[0]{}%
\providecommand \url  [0]{\begingroup\@sanitize@url \@url }%
\providecommand \@url [1]{\endgroup\@href {#1}{\urlprefix }}%
\providecommand \urlprefix  [0]{URL }%
\providecommand \Eprint [0]{\href }%
\providecommand \doibase [0]{http://dx.doi.org/}%
\providecommand \selectlanguage [0]{\@gobble}%
\providecommand \bibinfo  [0]{\@secondoftwo}%
\providecommand \bibfield  [0]{\@secondoftwo}%
\providecommand \translation [1]{[#1]}%
\providecommand \BibitemOpen [0]{}%
\providecommand \bibitemStop [0]{}%
\providecommand \bibitemNoStop [0]{.\EOS\space}%
\providecommand \EOS [0]{\spacefactor3000\relax}%
\providecommand \BibitemShut  [1]{\csname bibitem#1\endcsname}%
\let\auto@bib@innerbib\@empty
%</preamble>
\end{thebibliography}%


\begin{thebibliography}{1}
\bibitem{Caurier} E. Caurier {\it et al}., Rev. of Mod. Phys. \textbf{77}, 427 (2005).
\bibitem{MBPT} M. Hjorth-Jensen, T. T. S. Kuo and E. Osnes, Phys. Rept. {\bf 261}, 125 (1995).
\bibitem{Otsu10} T. Otsuka {\it et al}., Phys.~Rev.~Lett.~{\bf 105}, 032501
(2010). 
\bibitem{Holt12} J.\ D.\ Holt, T.\ Otsuka, A.\ Schwenk and T.\ Suzuki, 
J.\ Phys.\ G {\bf 39}, 085111 (2012).
\bibitem{Doba07} J. Dobaczewski {\it et al}., Prog. Part. Nucl. Phys. \textbf{59}, 432 (2007).
\bibitem{Stan08} M. Stanoiu {\it et al}., Phys. Rev. C \textbf{78}, 034315 (2008).
\bibitem{Camp06} C.~M.~Campbell {\it et al}., Phys. Rev. Lett. \textbf{97}, 112501 (2006).
\bibitem{Alex} A.~Lepailleur {\it et al}., Phys. Rev. Lett. \textbf{110}, 082502 (2013).
\bibitem{Hamamoto2012} I.~Hamamoto, Phys.~Rev.~C \textbf{85}, 064329 (2012).
\bibitem{Hagen2012b} G.~Hagen {\it et al}., Phys. Rev. Lett. \textbf{109}, 032502 (2012).
\bibitem{mueller} A.~C.~Mueller {\it et al}., Nucl. Phys. A \textbf{513}, 1 (1990). 
\bibitem{reed} A.~T.~Reed {\it et al}., Phys. Rev. C \textbf{60}, 024311 (1999).
\bibitem{USDA} B. A. Brown and W. A. Richter, Phys. Rev. C \textbf{74}, 034315 (2006).
\bibitem{Bogn14} S.\ K.\ Bogner {\it et al}., Phys. Rev. Lett. \textbf{113}, 142501 (2014).
\bibitem{Anne} R.~Anne {\it et al}., Nucl. Instr. Methods A \textbf{257}, 215 (1987). 
\bibitem{EXO} J. Simpson {\it et al}., Acta Physica Hungarica, New Series, Heavy Ion
Physics \textbf{11}, 159 (2000).
\bibitem{28Ne} V.~Tripathi {\it et al}., Priv. Communications (2011).
\bibitem{alex2} A.~Lepailleur, PhD Thesis, Universit\'e de Caen Basse Normandie, https://tel.archives-ouvertes.fr/tel-01057890/document
\bibitem{sissi} E.~Baron, J.~Gillet and M.~Ozille, Nucl. Instr. Methods A \textbf{362},
90 (1995).
\bibitem{dufour} J.~P.~Dufour {\it et al}., Nucl. Instr. Methods A \textbf{248}, 267 (1986).
\bibitem{SPEG} L.~Bianchi {\it et al}., Nucl. Instr. Methods A \textbf{276}, 509 (1999).
\bibitem{thesis} M.~Stanoiu, PhD thesis, Universit\'e de Caen (2002).
\bibitem{Vajta} Z. Vajta et al., Physical Review C \textbf{89}, 054323 (2014).
\bibitem{USDAm} W.~A.~Richter, S.~Mkhize and B.~A. Brown, Phys. Rev. C \textbf{78},
064302 (2008).
\bibitem{skx} B.~A.~Brown, Phys. Rev. C \textbf{58}, 220 (1998).
\bibitem{Holt13a} J.\ D.\ Holt, J.\ Men\'{e}ndez and A.\ Schwenk, 
Eur.\ Phys.\ J.\ A {\bf 49}, 39 (2013).
\bibitem{Holt13b} J.\ D.\ Holt, J.\ Men\'{e}ndez and A.\ Schwenk, 
Phys.\ Rev.\ Lett.\ {\bf 110}, 022502 (2013).
\bibitem{Caes13} C.\ Caesar {\it et al}. (R3B/LAND collaboration), 
Phys.\ Rev.\ C {\bf 88}, 034313 (2013).
\bibitem{Gallant} A. T. Gallant {\it et al}., Phys. Rev. Lett \textbf{113}, 082501 (2014).
\bibitem{Tsuk11} K.\ Tsukiyama, S.\ K.\ Bogner and A.\ Schwenk, 
Phys.\ Rev.\ Lett.\ {\bf 106}, 222502 (2011).
\bibitem{Tsuk12} K.\ Tsukiyama, S.\ K.\ Bogner and A.\ Schwenk, 
Phys.\ Rev.\ C {\bf 85}, 061304(R) (2012).
\bibitem{Hergert13} H.\ Hergert {\it et al}., Phys. Rev. C \textbf{87}, 034307 (2013).
\bibitem{CCOx} G.\ Hagen, M.\ Hjorth-Jensen, G.\ R.\ Jansen, R.\ Machleidt, and T.\ Papenbrock, 
Phys.\ Rev.\ Lett.\ {\bf 108}, 242501 (2012).
 \bibitem{Herg13} H.\ Hergert, S.\ Binder, A.\ Calci, J.\ Langhammer and R.\ Roth, 
 Phys.\ Rev.\ Lett.\ {\bf 110}, 242501 (2013).
\bibitem{Cipo13} A.\ Cipollone, C.\ Barbieri and P.\ Navr\'{a}til, 
Phys.\ Rev.\ Lett.\ {\bf 111}, 062501 (2013).
\bibitem{CCGT} A.\ Ekstr\"{o}m et al., arXiv:1406.4696
 \bibitem{EpelRMP} E.\ Epelbaum, H.-W.\ Hammer and U.-G.\ Mei{\ss}ner, 
 Rev.\ Mod.\ Phys.\ {\bf 81}, 1773 (2009).
\bibitem{EMPR} R. Machleidt and D. R. Entem, Phys. Rep. \textbf{503}, 1 (2011).
\bibitem {EM500} D. R. Entem and R. Machleidt, Phys. Rev. C {\bf 68}, 041001 (2003).
\bibitem{Roth3N400} R. Roth, S. Binder, K. Vobig, A. Calci, J. Langhammer and P. Navr\'atil, 
Phys. Rev. Lett. {\bf 109}, 052501 (2012).
\bibitem{BognSRG} S. K. Bogner, R. J. Furnstahl and R. J. Perry, 
Phys. Rev. C {\bf 75}, 061001(R) (2007).
\bibitem{CCNO3N} G. Hagen {\it et al}., Phys. Rev. C {\bf 76}, 034302 (2007).
\bibitem{RothNO}  R. Roth {\it et al}., Phys. Rev. Lett. \textbf{109}, 052501 (2012).




\end{thebibliography}
\end{document}